\documentclass[aps,prl,superscriptaddress,reprint,aps,showpacs]{revtex4-1}

\usepackage{graphicx}
\usepackage{dcolumn}
\usepackage{bm}
\usepackage{color}
\usepackage{ulem}
\usepackage{hyperref}
\usepackage{cleveref}

\newcommand{\FC}[1]{$F_{C}$ }
\newcommand{\TC}[1]{$T_{C}$ }

\begin{document}
\onecolumngrid
with kind permission from the American Physical Society, published as \href{https://journals.aps.org/prl/abstract/10.1103/PhysRevLett.119.086401}{Phys. Rev. Lett. 119, 086401 (2017)}
\\\\
\twocolumngrid
\title{Ultrafast Electronic Band Gap Control in an Excitonic Insulator}

\author{Selene Mor}
\affiliation{Fritz-Haber-Institut der MPG, Faradayweg 4-6, 14195 Berlin, Germany}

\author{Marc Herzog}
\affiliation{Fritz-Haber-Institut der MPG, Faradayweg 4-6, 14195 Berlin, Germany}\affiliation{Institute for Physics and Astronomy, University of Potsdam, Karl-Liebknecht-Str. 24-25, 14476 Potsdam, Germany}

\author{Denis Gole\v z}
\affiliation{Department of Physics, University of Fribourg, 1700 Fribourg, Switzerland}

\author{Philipp Werner}
\affiliation{Department of Physics, University of Fribourg, 1700 Fribourg, Switzerland}

\author{Martin Eckstein}
\affiliation{Department of Physics, University of Erlangen-N\"{u}rnberg, Staudtstrasse 7-B2, 91058 Erlangen, Germany}

\author{Naoyuki Katayama}
\affiliation{Department of Physical Science and Engineering, Nagoya University, 464-8603 Nagoya, Japan}

\author{Minoru Nohara}
\affiliation{Research Institute for Interdisciplinary Science, Okayama University, Okayama 700-8530, Japan}

\author{Hide Takagi}
\affiliation{Max Planck Institute for Solid State Research, 70569 Stuttgart, Germany}
\affiliation{Department of Physics, University of Tokyo, 113-8654 Tokyo, Japan}

\author{Takashi Mizokawa}
\affiliation{Department of Applied Physics, Waseda University, 169-8555 Tokyo, Japan}

\author{Claude Monney}
\email{monney@physik.uzh.ch}
\affiliation{Institute of Physics, University of Zurich, Winterthurerstrasse 190, 8057 Zurich, Switzerland}

\author{Julia St\"{a}hler}
\affiliation{Fritz-Haber-Institut der MPG, Faradayweg 4-6, 14195 Berlin, Germany}

\date{\today}

\begin{abstract} 
We report on the nonequilibrium dynamics of the electronic structure of the layered semiconductor Ta$_2$NiSe$_5$ investigated by time- and angle-resolved photoelectron spectroscopy. We show that below the critical excitation density of $F_{C} = 0.2$~mJ~cm$^{-2}$, the band gap \textit{narrows} transiently, while it is \textit{enhanced} above $F_{C}$. Hartree-Fock calculations reveal that this effect can be explained by the presence of the low-temperature excitonic insulator phase of Ta$_2$NiSe$_5$, whose order parameter is connected to the gap size.
This work demonstrates the ability to manipulate the band gap of Ta$_2$NiSe$_5$ with light on the femtosecond time scale.
\end{abstract}


\maketitle
Semiconductor materials are intensively studied due to their technological importance, for instance in the field of photovoltaics or data processing. 
The free carriers and charge redistributions generated across the band gap, for instance by light absorption, modify the screening properties and  electrostatic (Hartree) energies, which may lead to a shifting of the conduction (CB) and valence band (VB) toward each other and hence a narrowing of the band gap. This process is referred to as band gap renormalization \cite{Berggren,Oschlies,Pagliara,Chernikov,Pogna,Dou1997,Wagn1985,Wegk2014}. In addition, due to the attractive Coulomb interaction (CIA) between electrons and holes, bound states, called excitons, may form after electrons and holes relax toward the band extrema. 

\begin{center}
\begin{figure}[t!]
\includegraphics[width=0.80\columnwidth]{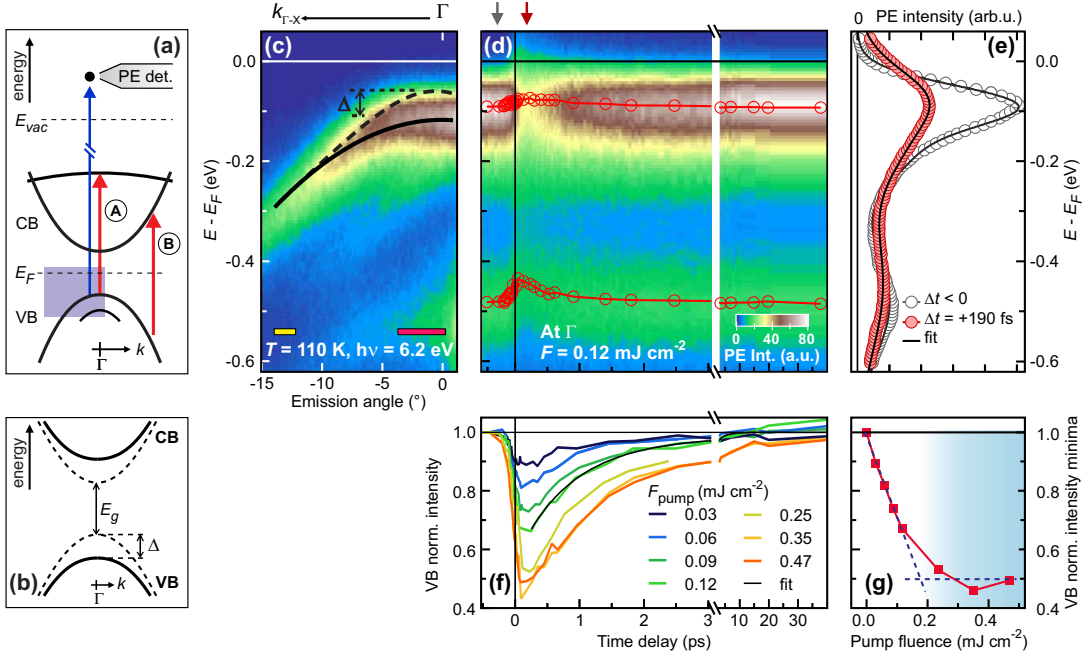}
\caption{(a) Schematic band structure of TNS at low temperature (see main text for details).
(b) Schematic EI (solid) and semiconductor (dashed) band structures. $E_g$ is the semiconductor band gap and $\Delta$ the enhancement of the band gap due to the transition to the EI phase. (c) PE spectrum of the TNS occupied electronic structure around $\Gamma$ at 110~K.  
The red and yellow bars indicate the momentum intervals of integration for the extracted EDCs at $\Gamma$ and at $k\neq0$, respectively. (d) PE intensity at $\Gamma$ as a function of electron binding energy with respect to $E_F$ (left axis) and pump-probe time delay (bottom axis) at 110~K. Markers indicate the transient energetic position of the two VBs. (e) EDCs at equilibrium (empty grey circles) and  190 fs after photoexcitation (solid red circles). (f) Transient population of the upper VB at $\Gamma$. (g) Minima from (f) as a function of pump fluence: 
Above $F_{C}$ $\approx$ 0.2 mJ cm$^{-2}$, the VB depopulation saturates at $\sim$50$\%$. Dashed lines are linear fits.}
\label{fig:1}
\end{figure}
\end{center}

When, for example in small gap semiconductors with low dielectric constant, the exciton binding energy $E_B$ exceeds the band gap $E_g$, the formation of excitons can occur without any external stimulus, resulting in a total energy gain. A macroscopic condensation of excitons has been predicted to occur in small gap semiconductors and to stabilize in a new ground state, the excitonic insulator (EI) phase \cite{Jerome,Keldysh,ZenkerBCSBEC}. This EI phase is characterized by a wider band gap $E_g$ + 2$\Delta$, where $\Delta$ is the energy difference between the parabolic (semiconductor) and flat upper (excitonic insulator) VB maxima [see Fig. \ref{fig:1}(b)], which is related to the exciton condensate density and hence to the order parameter of the phase transition \cite{Jerome}. 
A few materials have been suggested to exhibit a \textit{ground state} EI phase, like Ta$_2$NiSe$_5$ (TNS) \cite{Wakisaka_PRL,Seki}, TiSe$_2$ \cite{Cercellier, Monney_PRB} , and TmSe$_{0.45}$Te$_{0.55}$ \cite{Bucher, Bronold}. 

The quasi-one-dimensional TNS shows a second-order phase transition at a critical temperature of $T_{C} = 328$ K, accompanied by a structural distortion \cite{DiSalvo,Wakisaka_PRL}. Above $T_{C}$, TNS exhibits a direct band gap at $\Gamma$, which increases up to approximatively 0.3 eV with decreasing the temperature below $T_{C}$ \cite{OkamuraPrivate,LuNatureComm,Wakisaka_JSNM}. Simultaneously, an anomalous flattening and broadening of the VB occurs which has been interpreted as a signature of the formation of an EI phase in TNS \cite{Wakisaka_PRL,Wakisaka_JSNM,Seki,Kaneko}. Contrary to the indirect band gap material TiSe$_2$ \cite{RohwerTiSe,PorerTiSe}, the EI phase of TNS is not coupled to a charge density wave \cite{ZenkerEPC,KanekoHundEPC}.

In this work, we study the ultrafast, nonequilibrium dynamics of the occupied electronic structure of TNS using time- and angle-resolved photoelectron spectroscopy (trARPES).  We show that, at modest near-infrared (NIR) excitation densities ($<$~$F_{C} = 0.2$~mJ~cm$^{-2}$), the occupied electronic structure exhibits an abrupt shift towards the Fermi energy, $E_F$. These dynamics are indicative of a transient \textit{narrowing} of the band gap induced by free-carrier screening of the CIA and Hartree shifts. However, above \FC, the direct band gap of TNS is transiently \textit{enhanced} on the time scale of 200 fs. As this behavior is opposite to that of ordinary semiconductors, we argue, on the basis of Hartree-Fock calculations, that it is a direct consequence of the transient enhancement of the order parameter of the exciton condensate. We demonstrate that we can either increase or decrease the size of the band gap in TNS by tuning the NIR excitation density, the proof of principle of ultrafast electronic 
\onecolumngrid
\twocolumngrid 
\noindent band gap control in a semiconductor.

The electronic structure of TNS and the experimental scheme \cite{methods} are illustrated in Fig. \ref{fig:1}(a). From the upper VB, two electronic transitions can be optically excited by an ultrashort NIR laser pulse with a photon energy of $h\nu_{\mathrm{pump}}$ = 1.55 eV: (A) to the flat unoccupied \textit{d} band at $\Gamma$ (i.e., \textit{k} = 0), and (B) to the lowest CB, but at larger \textit{k} vectors. The occupied electronic band structure indicated by the blue-shaded region and its nonequilibrium dynamics are monitored by trARPES using $h\nu_{\mathrm{probe}}$ = 6.2 eV.
The energy resolution is 86~meV and the upper limit for the time resolution is 110~fs \cite{methods}.  

At $T=110$~K  and at equilibrium, i.e.,  without laser excitation, TNS exhibits, as shown in Fig. \ref{fig:1}(c), two VBs with maxima at $E-E_F \approx -0.5$ and $-0.1$ eV, respectively, which is consistent with previous angle-resolved photoelectron spectroscopy results \cite{Wakisaka_PRL}. Considering the optical gap of $0.3$ eV \cite{OkamuraPrivate,LuNatureComm}, the sample appears to be slightly $p$ doped \cite{methods}. Figure \ref{fig:1}(d) shows the temporal evolution of the photoelectron (PE) intensity at $\Gamma$ and $T$ = 110 K in false color after moderate photoexcitation \footnote{The energetic offset of the band positions in graphs \ref{fig:1}(c) and \ref{fig:1}(d) (negative delays), is attributed to the heat load of the pump pulse that is not fully dissipated before the next pump pulse arrives. This energy shift of 19~meV corresponds to a temperature increase of 70~K \cite{Wakisaka_JSNM}, i.e. leaving the sample below \TC.}. The data reveal a rapid response to the optical excitation at $\Delta t$ = 0~fs: a massive suppression of PE intensity within the first 500~fs, and an energy shift of both VBs towards $E_F$. This is further illustrated by two energy distribution curves (EDCs), before (empty gray circles) and 190 fs after (solid red circles) NIR photoexcitation, in Fig. \ref{fig:1}(e). 
For a quantitative analysis of the monitored electronic structure dynamics, we fit such EDCs at various pump-probe time delays \cite{methods} and obtain the transient amplitude of the spectral function and energetic position at $\Gamma$ for each band. Two exemplary fits are shown in Fig \ref{fig:1}(e) (solid lines). The resulting peak positions are depicted in panel (d) (red circles), confirming that both VBs show an abrupt upward shift upon photoexcitation, which is more pronounced for the lower band. It should be noted that the general trend of the peak shift is visible by eye in the normalized data and robust against the choice of fit function \cite{methods}.

Figure \ref{fig:1}(f) displays the temporal evolution of the flat top VB integrated intensity at $\Gamma$ for different incident pump fluences: the VB is increasingly depopulated by the increasing excitation density and its occupation recovers, on average, within 870 $\pm$ 210~fs via interband carrier thermalization, likely assisted by scattering with phonons \footnote{The change of the integral of the VB spectral function can be interpreted as a measure for its occupation evolution under the assumption of temporally constant transfer matrix elements.}.
The population minima scale linearly with the pump fluence up to a critical value $F_{C}$ $\approx$~0.2~mJ~cm$^{-2}$ above which the VB depopulation saturates at $\sim$ 50$\%$, as shown in Fig. \ref{fig:1}(g). It seems likely that this behaviour results from an absorption saturation \cite{Einstein1916}. We also observe this saturation effect in time-resolved reflectivity measurements \cite{HerzogOptics}.

\begin{figure}
\includegraphics[width=0.80\columnwidth]{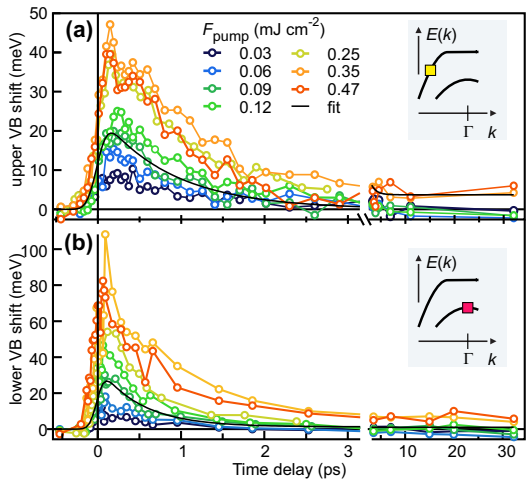}
\caption{(a) Shift of the upper VB at $k \neq$ 0 as a function of pump-probe time delay for different excitation densities. (b) Same analysis for the lower VB at $\Gamma$. Black lines in (a) and (b) are single exponential fits to the data convolved with the envelope of the laser pulses.}
\label{fig:2}
\end{figure}

In Fig. \ref{fig:2}, we focus on the shifts observed for the upper VB at $k\neq 0$ and the lower VB at $k=0$, which are expected to remain largely unaffected by photoinduced changes to the exciton condensate. Different trace colors correspond to different excitation densities. In both cases, the bands shift abruptly toward lower binding energy, i.e. toward $E_F$. We quantify these dynamics by fitting a single exponential decay convoluted with the envelope of the laser pulses \cite{methods} as exemplarily shown by the black curves in Fig. \ref{fig:2}. At all fluences and for both VBs the fits yield comparable recovery times (of about 1~ps). This is in good agreement with the time scale observed for the population dynamics of the photoinduced carriers at $\Gamma$, see Fig. \ref{fig:1}(f). Moreover, the band shift becomes stronger with increasing excitation density up to the same critical fluence \FC at which the absorption saturation of the flat top VB occurs. Since we do not observe a rigid spectrum shift, nor changes of the low-energy cutoff \cite{methods}, the measured shifts and saturation cannot be due to surface photovoltage effects. Thus, we interpret the upward shift as the fingerprint of a \textit{transient shrinking} of the band gap following the photoexcitation. Our findings strongly suggest that the abrupt band gap narrowing is a consequence of the photocarriers generated by photoexcitation (A) from the flat top VB.

\begin{figure}
\includegraphics[width=0.80\columnwidth]{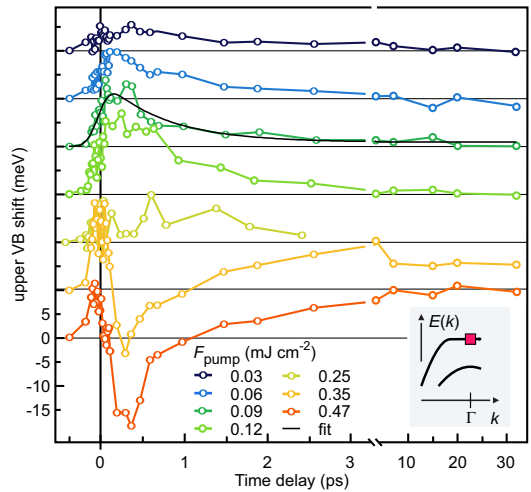}
\caption{Shift of the upper VB at $\Gamma$ as a function of pump-probe time delay for different excitation densities.}
\label{fig:3}
\end{figure}

The photoinduced shift of the flat top VB at $\Gamma$ is shown in Fig. \ref{fig:3}. For $F < F_{C}$ (blue to light green traces), the dynamics resemble the trend observed at $k\neq 0$ [see Fig. \ref{fig:2}(a)], although less pronounced at the respective fluences. The same fit function (solid black) as in Fig. \ref{fig:2} is used to quantify these early dynamics. However, for larger fluences (orange and red traces), we observe a shift in the \textit{opposite} direction, delayed by $\sim$ 200 fs, such that the flat top VB maximum transiently lies at \textit{higher} binding energies than its equilibrium position.  After approximately 1 ps, this effect relaxes by an upward shift of the band that even ``overshoots'' the equilibrium position. The delayed downshift of the flat top VB is sufficiently pronounced to be directly seen in the raw PE data \cite{methods}. These results demonstrate that the band gap of TNS is transiently \textit{enhanced} by strong photoexcitation.

\begin{figure}
\includegraphics[width=0.80\columnwidth]{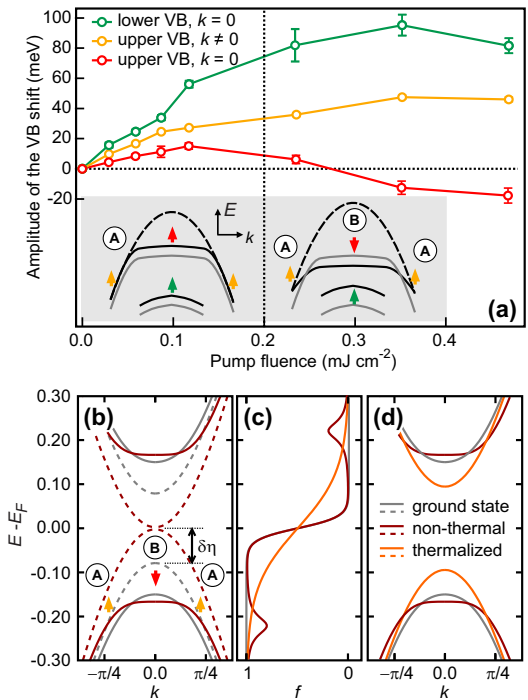}
\caption{(a) The fit amplitude of the shift of the lower VB at $\Gamma$ (green) and of the upper VB at \textit{k} $\neq$ 0 (yellow) and at $\Gamma$ (red) as evaluated for a delay of 255~fs [see Fig. \ref{fig:3}] as a function of the incident pump fluence. In the inset, the band structure dynamics are schematized for two excitation regimes, below (left) and above (right) \FC, respectively. (b) Dispersion relation for the ground state (gray) and a nonthermal distribution function (dark red), obtained from the Hartree-Fock calculations. The dashed lines show the bare semiconducting dispersion (with Hartree shift included). (c) Distribution functions corresponding to the dispersion relations in panel (b) and (d). (d) Comparing the dispersions of the thermal state at elevated temperature (orange) with that of the ground state and the photoexcited nonthermal state.}
\label{fig:4}
\end{figure}

Figure \ref{fig:4}(a) summarizes the observed different VB shifts at three points in the TNS electronic structure as a function of the incident fluence. Clearly, the upward shift occurring in the upper VB at $k \neq$ 0 (yellow) and in the lower VB at $\Gamma$ (green) is enhanced with increasing excitation density and exhibits a slope change at the same critical fluence $F_{C}$  as the VB depopulation threshold shown in Fig. \ref{fig:1}(f). The transient shift of the flat top VB at $\Gamma$ follows, on the contrary, a nonmonotonic curve, revealing that two competing phenomena are at play: While, for \FC, the flat top VB qualitatively follows the trend of the yellow and green curves, a \textit{time-delayed} process counteracts this band gap narrowing for $F>$ \FC. This is reflected in a shift away from $E_F$ with respect to the equilibrium position  and, consequently, a transient widening of the band gap at $\Gamma$. 

Both effects are illustrated by the insets in Fig. \ref{fig:4}(a). As the upward shift saturates at $F_{C}$, we conclude that it must be connected to the transient free carrier density that is generated by excitation (A) [cf. Fig. \ref{fig:1}(a)] where carriers from the flat top VB are transferred to the upper CB, a process that saturates at a population redistribution of $\sim$50~$\%$. The upward shift is, thus, driven by a band gap renormalization that is caused by transiently enhanced free carrier screening, as expected for a semiconductor under optical excitation \cite{Berggren,Oschlies,Pagliara,Chernikov,Pogna,Dou1997,Wagn1985,Wegk2014}. The extraordinary behavior of the \textit{downward} shift of the flat top VB is analyzed with regard to the presence of an exciton condensate in the following.

In particular we show that, in the presence of an exciton condensate, nonthermal carrier distributions and an electrostatic shift of the underlying band structure can lead to an enhancement of the condensate density and a corresponding  enhancement of the band gap at $\Gamma$.
To demonstrate this, we perform a Hartree-Fock calculation for the simplified case of a one-dimensional two-band system, representing the upper VB and lower CB \footnote{TNS is a highly anisotropic material, where the dispersion perpendicular to the chain is about 10 times smaller than along the chains \cite{Wakisaka_JSNM}, which allows us to model the system as a one-dimensional chain.}. 
We emphasize that the mechanism proposed in the following does not depend on the dimensionality of the system, and its effect would most likely be further enhanced by including higher lying orbitals in the model, due to additional Hartree shifts.
The bands are shown in Fig. \ref{fig:4}(b), where the dashed (solid) curves represent the noninteracting (interacting) case. 
We now distinguish two essential effects resulting from the photoexcitation.
(i) The formation of a nonthermal carrier distribution, which is \textit{explicitly} included in the calculation as shown in Fig.~\ref{fig:4}(c) (dark red). This effect leads to an electrostatic reduction of the splitting between the bare semiconductor VB and CB.
(ii) The generation of additional free charge carriers that involves also other bands in the experiment and enhances the screening in the semiconductor.
We mimic this effect by an additional reduction of the bare band splitting. The total shift of the bare dispersion is denoted by $\delta\eta$ in Fig. \ref{fig:4}(b).
Clearly, we observe that the interplay of (i) and (ii) in the interacting case can lead to an enhancement of the band gap at $\Gamma$ while the bands shift towards each other at larger $k$ [compare solid dark red and gray lines in Fig. \ref{fig:4}(b)] as observed in the experiment.

The Hartree-Fock calculations suggest that the enhancement of the band gap is related to an increase of the exciton condensate density (and, hence, of the order parameter of the EI phase), which relies on two effects. On the one hand, one finds that for a given value of the CIA the equilibrium condensate density increases if the (noninteracting) conduction and valence bands are shifted closer together, because then excitons are formed from more resonant states. The electrostatic effect of a photoexcited electron density $n_{\mathrm{ex}}$ (i.e., the Hartree contribution to the interaction) can produce such a shift, and thus tends to enhance the density of condensed pairs \footnote{We define $n_{\mathrm{ex}}=n_{\mathrm{ex}}^{\mathrm{nonth}}-n_{\mathrm{ex}}^{\mathrm{ini}}$ as the difference of the integrated weight for the occupied density of states above the Fermi level between the nonthermal $n_{\mathrm{ex}}^{\mathrm{nonth}}$  and the initial state $n_{\mathrm{ex}}^{\mathrm{ini}}$.}. On the other hand, a large $n_{\mathrm{ex}}$ weakens the condensate, similar to a thermal population of quasiparticle states. The crucial observation is that this potentially detrimental effect of the photoexcited population on the order parameter can be much weaker in a nonthermal state than in a thermal one: for the parameters of Fig.~\ref{fig:4}(b) and an excitation density of $n_{\mathrm{ex}}=0.02$, we use the nonthermal carrier distribution displayed in panel (c) (dark red). In this situation the exciton condensate is enhanced. The effect is robust for a broad range of nonthermal distribution functions \cite{methods}, while in the fully thermalized state [solid orange in Fig. 4(d)] the temperature would be above the melting temperature of the condensate. Therefore, as thermalization occurs, we expect a band gap reduction, as indeed observed in the experiments after about 1 ps [cf. Fig. \ref{fig:3}].

In conclusion, the present experimental and theoretical work reports and analyzes a transient band gap enhancement in the layered semiconductor Ta$_2$NiSe$_5$ upon photoexcitation due to the existence of an excitonic insulator phase in this material. We propose that the strengthening of the condensate in the perturbed state is a consequence of the low-energy electronic structure of this material, which provides different channels for resonant excitation and the possibility of significant Hartree shifts and nonthermal quasiparticle distributions. A critical photoexcitation density is observed, which separates regimes in which the band gap can be either increased or decreased. This work not only provides new insights about the \textit{nonequilibrium} properties of excitonic insulators but also proves the possibility of controlling the band gap of such materials by tuning the photoexcitation fluence.

D.G. and P.W. acknowledge support from ERC starting Grant No. 278023, ERC Consolidator Grant No. 724103 and from SNSF Grant No. 200021-140648. C.M. acknowledges the support by the SNSF grant No. $PZ00P2\_ 154867$.

\bibliography{references}
\newpage
\noindent
SUPPLEMENTAL MATERIAL TO ``ULTRAFAST ELECTRONIC BAND GAP CONTROL IN AN EXCITONIC INSULATOR''
\subsection{Experimental details}
TNS single crystalline samples were prepared by reacting the elementals nickel, tantalum and selenium with a small amount of iodine in a evacuated quartz tube. The tube was slowly heated and kept with a temperature gradient from 950$^{\circ}$C to 850$^{\circ}$C for 7 days, followed by slow cooling. Single crystalline samples with a typical size of 0.04~x~1~x~10 mm$^3$ were obtained in the cooler end. 

For trARPES, the samples were cleaved \textit{in situ} at room temperature under ultrahigh vacuum conditions (1.6~x~10$^{-10}$ mbar) and then slowly cooled down to 110~K using liquid N$_2$. TNS was optically excited by the p-polarized, fundamental output (h$\nu_{\mathrm{pump}}$ = 1.55~eV) of a regeneratively amplified Ti:Sa laser system working at a repetition rate of 40 kHz. The photoinduced changes to the electronic structure along the Ni chain direction of TNS were probed by time-delayed p-polarized probe pulses (h$\nu_{\mathrm{probe}}$ = 6.2~eV), generated by frequency quadrupling of the fundamental. The two-photon photoemission signal of TNS at high kinetic energies represents a cross correlation of the Gaussian pump and probe laser pulses and was used to estimate an upper limit for the time resolution of 110~fs. The photoelectrons were detected by a  hemispherical analyzer (SPECS Phoibos 100) held at a bias voltage of 0.5 V with respect to the sample holder. The energy resolution of 86~meV is obtained as root sum squared of the UV pulse bandwidth (25 meV) and the instrument resolution (82~meV) estimated \textit{in situ} from the low-energy secondary electron cut-off of a direct photoemission spectrum of the metallic sample holder. Due to the high statistics of our raw data, it was possible to resolve \textit{relative} energy shifts of only few meV. PE spectra are plotted as a function of energy with respect to the equilibrium Fermi level $E-E_{\mathrm{F}}=E_{\mathrm{kin}}-$h$\nu_{\mathrm{probe}}+\Phi$. $\Phi$ is the work function. $E_{\mathrm{F}}$ was determined independently on the metallic sample holder, which was in direct electrical contact with the sample. We can neglect sample charging as the origin of the observed transient spectral shifts as (i) no \textit{rigid} shift of the angle-resolved PE spectra was observed (parts of the spectrum shift up, others down) and (ii) the low-energy secondary electron cut-off remains unchanged.
This is exemplified in Fig. \ref{fig:1SM}, where we compare EDCs at $\Gamma$ showing the whole ARPES spectra measured before $\Delta t = 0$ fs and at $\Delta t = 185$ fs (for a fluence of 0.12~mJ/cm$^2$) at 110 K. While the peaks of the lower and upper VB are transiently modified by the pump pulse, this shows that the low-energy cut-off (at a binding energy of about -0.8~eV) remains unchanged. This has been checked for different fluences and at different time delays.
\renewcommand{\thefigure}{SM\arabic{figure}}
\setcounter{figure}{0}
\begin{center}
\begin{figure}[h]
\includegraphics[width=0.80\columnwidth]{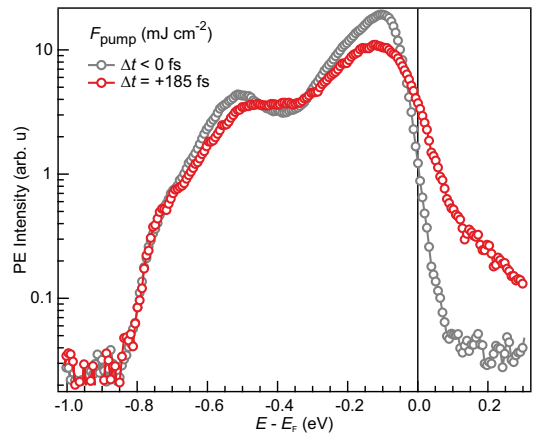}
\caption{EDCs at $\Gamma$ before (empty grey) and at 185 fs (empty red) after photoexcitation for a fluence of 0.12 mJ/cm$^2$.}
\label{fig:1SM}
\end{figure}
\end{center}

\subsection{Experimental data analysis}

\begin{center}
\begin{figure}[h]
\includegraphics[width=0.80\columnwidth]{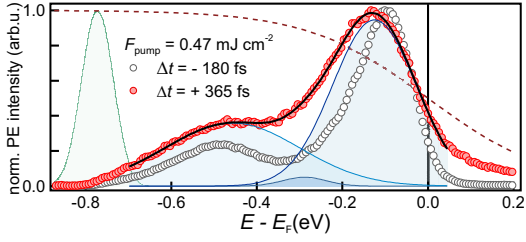}
\caption{Normalized EDCs at $\Gamma$ before $\Delta t = 0$ fs (empty grey) and at $\Delta t = 365$ fs after (solid red) photoexcitation with 0.47 mJ/cm$^2$ at 110~K. 
The exemplary fit curve (black line) is composed of three Gaussian peaks (blue) multiplied by a Fermi-Dirac occupation distribution (red dashed), and convolved with a Gaussian accounting for the energy resolution (green).}
\label{fig:2SM}
\end{figure}
\end{center}

EDCs were fitted with a sum of three Gaussian peaks multiplied by the Fermi-Dirac occupation distribution (FDD) and convoluted with another Gaussian to account for the energy resolution as exemplarily shown in Fig. \ref{fig:2SM}. The Gaussian peak at higher binding energy fits to the lower VB, while the sum of the other two Gaussians is used to reproduce the asymmetric shape of the upper VB.
The EDC at $\Delta t = 365$ fs (solid red markers) in Fig. \ref{fig:2SM} clearly shows the downward shift of the upper flat top VB upon strong photoexcitation. This delayed downshift is sufficiently pronounced to be directly seen in the raw PE data, as evidenced by the normalized EDCs for different time delays before (empty grey) and after (solid red) photoexcitation. 
The time-dependent energetic position of the intensity maximum of the combined peak was used as a measure for the transient shift of the upper VB\footnote{The separate shifts of B and C exhibit the same qualitative behavior (not shown).}.
Since the energy shift of both peaks is sufficiently strong to be seen by eye in the raw data, different fit functions (not shown, e.g. two Gaussians w/o FDD and three Gaussians w/ FDD) yield qualitatively the same result. 
Tests demonstrate that apparent band shifts due to FDD broadening can only account for a maximum downward shift of 5~meV, far below the observed maximum shift of 18~meV.

The transient population of the VBs has been estimated from the area under the corresponding peak at each pump-probe delay.
It should be noted that the electric field of the intense pump pulses might affect the spectral shape probed by ARPES. However, such effects occur only during the pump pulse duration \cite{MahmoodField} and could solely have an impact on our results at time delays $\pm$~50~fs.

The band gap of TNS is approximately 0.3 eV \cite{OkamuraPrivate}. In Ref. \cite{LuNatureComm}, this corresponds to the isobestic point in the optical data. From our ARPES data at equilibrium (without pump photons) and at 110 K, we find that the upper VB lies at $E-E_{\mathrm{F}}$ = -0.11 eV, which is smaller than half the band gap value. This suggests a slight p-doping of our sample.

\begin{center}
\begin{figure}[h]
\includegraphics[width=0.80\columnwidth]{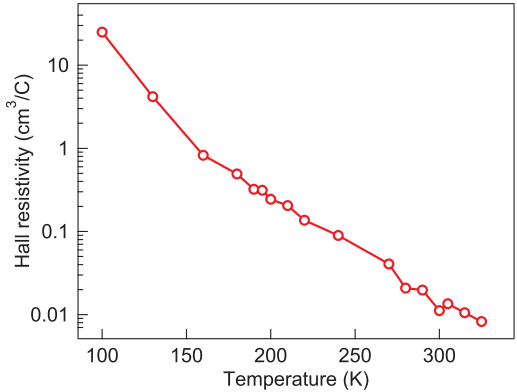}
\caption{Hall resistivity of our TNS samples.}
\label{fig:2bSM}
\end{figure}
\end{center}

This assertion about the p-doping of our sample is further supported by a Hall resistivity measurement performed on another sample from the same growth batch (see Fig. \ref{fig:2bSM}), which confirms that the Hall resistivity is hole-like (positive) on the whole temperature range displayed here.

\subsection{Theoretical modeling}
We investigate the nonthermal state of the EI by considering a one-dimensional two-band system of spin-less fermions (spin-degrees of freedom are not included) with direct band gap, 
 $H=H_{0}+H_{\text{int}},$ 
where the noninteracting part is
$$H_{0}=\sum_{k,\alpha} (\epsilon_{k,\alpha}+\eta_{\alpha})c_{k,\alpha}^{\dagger}c_{k,\alpha},$$
with the band dispersion $\epsilon_{k,1(2)}=-(+)2t_0 \cos(k)$. The $c_{k,\alpha}$ denote the annihilation operators for an electron with momentum $k$ in orbital $\alpha=1,2.$ The 
bare band splittings $\eta_{1,2}$ are chosen such that band $1\:(2)$ is totally occupied (unoccupied). The hopping parameter $t_0$ is chosen such that the ground state dispersion in the unordered state matches the experimental one.
For the interaction we consider a local density-density CIA of the form
$$H_{\text{int}} 
=
\frac{1}{2} \sum_i U n_{i,1} n_{i,2}.
$$
We determined the interaction $U$ by the comparison of calculated and equilibrium ARPES spectra in the ordered and unordered state. Explicitly, for the ground state calculations, we used $t_0=0.26$ eV, $U/t_0=3.0$ and the relative bare band splitting $(\eta_{2}-\eta_{1})/t_0=2.1$.

\begin{center}
\begin{figure}
\includegraphics[width=0.80\columnwidth]{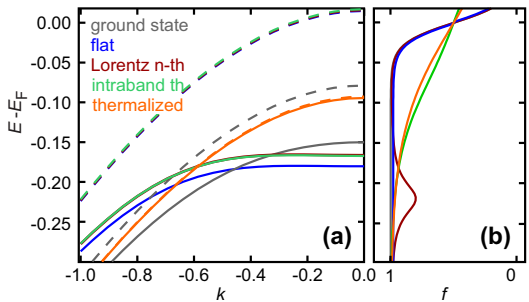}
\caption{(a) Comparison of the ground state dispersion relation to the ones obtained from different nonthermal distribution functions, namely in blue a constant population of holes (electrons), in dark red a dip (hump) in the Fermi-Dirac distribution function below (above) $E_{\mathrm{F}}$, respectively, in green the corresponding \textit{intra}band thermalized distribution at elevated temperatures. The negative energy parts of the corresponding distribution functions are presented in panel (b).}
\label{fig:5SM}
\end{figure}
\end{center}

The excitonic instability arises because of the attractive CIA between the electrons in the upper band and the holes in the lower band, leading to a condensation of the excitons which are formed across the direct band gap. 
The order parameter of the condensate is $\rho_{12}=\langle c_{k,1}^\dagger c_{k,2} \rangle \neq 0.$ In order to solve the problem we employ standard Hartree-Fock calculations based on the mean-field decoupling of the interaction term. The experimentally observed abrupt band gap narrowing is modeled by a reduction of the bare band splitting $\eta_2-\eta_1$ between the VB and CB. 
The effect of the nonthermal distribution on the band gap size $\Delta$ is determined from the self-consistent Hartree-Fock calculation. In order to show that the conclusions do not depend on the particular choice of the distribution function, we compared several parametrizations of the nonthermal distribution function, see Fig.~\ref{fig:5SM}(b): (i) additional constant population of holes (electrons), (ii) Lorentzian peak (dip) in the Fermi-Dirac distribution function below (above) $E_{\mathrm{F}}$, namely $f_{\text{nth}}(\epsilon,A,\overline E)=f_{\text{FD}}(\epsilon,0)+A\gamma^2/((\epsilon-\overline E)^2+\gamma^2),$ where $f_{\text{FD}}(\epsilon,\mu)$ is the Fermi-Dirac distribution, $A$ is the amplitude, $\gamma$ the width and $\overline E$ the center position of the Lorentzian peak (dip), (iii) intraband thermalized distributions at elevated temperatures.\\
As shown in Fig. \ref{fig:5SM}(a), for all these nonthermal distributions with the same excitation density $n_{ex}$ the band gap at $\Gamma$ is enhanced except for the \textit{fully} thermalized one (solid orange in Fig. \ref{fig:5SM}(e)). This means that photoexcitation of an EI can result in a transient enhancement of the order parameter, if thermalization is sufficiently delayed by, for instance, the presence of the band gap, consistently with the experimental observations at later time 
delay. 

The theory that we present does not aim at realistically reproducing the experimental setup, including the photoexcitation and all possible relaxation processes. We rely on the separation of the time scales for the relaxation, and assume that the intraband electron-electron scattering leads to a fast relaxation and thermalization within each band, while the thermalization between the bands is mediated by electron-phonon interactions whose characteristic time is much longer. This leads to a nonthermal distribution function which exists in a relatively long time-window. The mean-field analysis is well suited to describe the state of the system, i.e., its band structure and order parameter, in the presence of such a non-thermal electron distribution. In order to simulate the relaxation and thermalization and therefore predict the precise time-dependent form of the nonthermal distribution function, however, one needs to include processes like electron-electron and electron-phonon scattering in the real-time description, which can be obtained by more advanced descriptions beyond mean-field, such as GW or DMFT. In these approaches, however, the timescales that one can reach are limited by the large cost in memory and computer time. Therefore, there is a need to develop new techniques to describe the long time behavior that includes the relaxation and thermalization dynamics, which is a challenging problem for future theoretical studies.


\end{document}